\documentclass[twocolumn]{aastex62}
\usepackage{amsmath}
\usepackage{float}

\graphicspath{{./}{figures/}}

\shorttitle{Possible Transfer of Life by Earth-Grazing Objects to Exoplanetary Systems}

\begin{document}

\title{Possible Transfer of Life by Earth-Grazing Objects to Exoplanetary Systems}

\email{amir.siraj@cfa.harvard.edu, aloeb@cfa.harvard.edu}

\author{Amir Siraj}
\affil{Department of Astronomy, Harvard University, 60 Garden Street, Cambridge, MA 02138, USA}

\author{Abraham Loeb}
\affiliation{Department of Astronomy, Harvard University, 60 Garden Street, Cambridge, MA 02138, USA}



\begin{abstract}
Recently, a 30-cm object was discovered to graze the Earth's atmosphere and shift into a Jupiter-crossing orbit. We use the related survey parameters to calibrate the total number of such objects. The number of objects that could have exported terrestrial microbes out of the Solar System is in the range $2 \times 10^{9}$--$3 \times 10^{11}$. We find that $10^{7}$--$10^{9}$ such objects could have been captured by binary star systems over the lifetime of the Solar System. Adopting the fiducial assumption that one polyextremophile colony is picked up by each object, the total number of objects carrying living colonies on them upon capture could be $10$--$10^3$.

\end{abstract}



\section{Introduction}
Panspermia is the conjecture that life can propagate from one planet to another \cite{Wesson2010, Wickramasinghe2010}. One~version of panspermia involves Solar System bodies grazing the Earth's atmosphere, picking up microbes, and~being ejected from the Solar System. About $\sim$1--60 long-period comets and interstellar objects could have to have undergone such a process over the lifetime of the Solar System~\cite{Siraj2019} .~It has also been shown that ejected objects can be gravitationally captured by other star~systems \cite{Ginsburg2018,Lingam2018}.

Recently,~\citet{Shober2019} reported a detection by the Desert Fireball Network (DFN) (\url{http://fireballsinthesky.com.au/}) of a $\sim$30-$\mathrm{cm}$ object that reached a minimum altitude $\sim$58.5 $\; \mathrm{km}$ in the Earth's atmosphere during a $90 \; \mathrm{s}$ grazing event, and~was transferred from an Apollo-type orbit into a Jupiter-family comet orbit, making it likely to be ejected from the Solar System during a future gravitational encounter with Jupiter. This detection represents a new class of objects that can pick up life in the Earth's atmosphere before being ejected from the Solar System: rocky, inner-Solar System bodies that are scattered into Jupiter-crossing orbits after their grazing interactions with Earth and are subsequently ejected. These~objects are important, since, based on their characteristically higher densities relative to icy, outer-Solar System bodies, they can survive passes through the Earth's atmosphere at significantly smaller~sizes.

Here, we study the likelihood of life-bearing Solar System bodies being captured by exoplanetary systems. First, we discuss the collection of microbial life during the transporting body's passage through the atmosphere as well as the survival lifetime in space. Next, we use the detection be \citet{Shober2019}  to calibrate the number of similar objects that could have picked up life in the Earth's atmosphere and were likely ejected from the Solar System. We then use a Monte Carlo simulation to estimate the total number of such objects that were captured by stellar binary systems over the lifetime of the Solar System, as~well as the number of these objects that likely harbor living microbes at the time of capture. Finally, we summarize our main~conclusions.


\section{Collection of Microbial~Life}

A 30-cm object grazing the atmosphere for 90 s at a speed of $15.5 \mathrm{\; km \; s^{-1}}$ should collect $\sim$$10^4$ microbial colonies during its trip through the atmosphere, based on Equation~(5) from~\cite{Siraj2019} for the number of colonies collected,
\begin{equation}
    \sim 10^4 \; \mathrm{colonies} \; \left( \frac{d}{30 \; \mathrm{cm}} \right)^2 \left( \frac{v}{15.5 \; \mathrm{km \; s^{-1}}} \right) \left( \frac{\tau}{90 \; \mathrm{s}} \right) \; \; ,
\end{equation}
where $d$ is the object diameter, $v$ is the speed, and~$\tau$ is the time spent in the~atmosphere.

It has been demonstrated that many microbes can likely survive the accelerations associated with transfer onto the grazing object \cite{Mastrapa2001, Deguchi2011}, although~specific detections in the atmosphere linked to microbes that have undergone such laboratory testing have not yet been~made.

Polyextremophiles such as \textit{Deinococcus radiodurans} are estimated to die on an exponential timescale of $\sim10^5 \; \mathrm{years}$ with minimal radiation shielding \cite{Ginsburg2018, Phillips2013, Burchell2004, Mileikowsky2000}. While the abundance of polyextremophiles in the atmosphere is yet unconstrained, in~our analysis, we assume a fiducial estimate of 1 polyextremophile colony ($\lesssim$10\textsuperscript{$-$4} of incident microbe colonies) per transporting object. The~interiors of objects with diameters greater than 20 cm are not expected to be heated to more than $100~^{\circ}$C during a passage through the atmosphere \cite{Napier2004}, and~many asteroids are known to have significant porosity \cite{Britt2000, Britt2006},  thus  we adopt a model in which incident microbes become lodged inside the object and shielded from the exterior heating. We note that this assumption is not critical; increasing the minimum altitude by a few scale heights above the observed altitude of the \citet{Shober2019} bolide would result in minimal heating (<100~\textsuperscript{$\circ$}C) while only reducing the atmospheric cross section by a small factor, potentially allowing for microbes on the surface in the case of non-porous~objects.

\section{Number of Ejected~Objects}

The DFN uses $\sim$50 cameras to cover $\sim$2.5 $\times 10^6 \; \mathrm{km^2}$ of the sky, or~$\sim$5 $\times 10^{-3}$ of the Earth's surface \cite{Devillepoix2016}. While the DFN only had four cameras in 2007, we conservatively estimate the relevant observation parameters for the~\citet{Shober2019} detection to be the result of a survey that lasted $\sim$10 years and \mbox{covered $\sim$5 $\times 10^{-3}$ of} the Earth's surface. This leads to a rate estimate of $\sim$20 $\; \mathrm{year^{-1}}$ of 30-cm objects penetrating the Earth's atmosphere and eventually likely being ejected from the Solar System, with~95\% Poisson bounds of $6 \times 10^{-1}$--$1.1 \times 10^2 \; \mathrm{year^{-1}}$. Over~$3 \times 10^9 \; \mathrm{year}$, this translates to $\sim$6 $\times 10^{10}$ ejected objects, with~95\% Poisson bounds of $2 \times 10^9$--$3 \times 10^{11}$.

We use Equation~(9) from~\cite{Collins2010},
\begin{equation}
    \log_{10} Y_{obj} = 2.107 + 0.0624 \sqrt{\rho_{obj}},
\end{equation}
to estimate the yield strength $Y_{obj}$ of the~\cite{Shober2019} meteor to be $\sim$6 $\times 10^5 \; \mathrm{Pa}$, given its density estimate of $\rho_{obj} \sim$3500 $\; \mathrm{kg \; m^{-3}}$. We then use Equation~(10) from~\cite{Collins2010},
\begin{equation}
    Y_{\mathrm{obj}} = \rho_{air}(z_{\star}) \; v^2(z_{\star}),
\end{equation}
where $\rho_{air}(z)$ is the density of air at altitude $z$, and~the known speed of $v = 15.5 \; \mathrm{km \; s^{-1}}$ to estimate that the altitude of breakup as $z_{\star} \sim$48 $\; \mathrm{km}$, which is $\sim 10 \; \mathrm{km}$ below the observed minimum altitude of the~\citet{Shober2019} meteor. Breakup is therefore not considered to be a significant factor in this analysis for typical rocky compositions, although~it is noted that the~\citet{Shober2019} meteor experienced a breakup event, which may be indicative of a composite~makeup.

The drag force on the object can be written as,
\begin{equation}
    \frac{d(mv)}{dt} = -\frac{1}{2} C_D \rho_{air}(z) v^2 A,
\end{equation}
where $C_D$ is the drag coefficient, taken to be unity for typical meteor speeds; $m$ and $v$ are the mass and speed of the object; and~$A$ is the cross-sectional area of the object. We adopt the limit of constant mass, assume $v \approx 15.5 \mathrm{\; km \; s^{-1}}$, and~require that the maximum change in speed is $-4 \mathrm{\; km \; s^{-1}}$, so that the object will possess more than the escape speed from the Earth and not fall to the ground. Over~90 s of travel in the atmosphere, the~latter constraint can be expressed as a maximum acceleration, $a_{max} \lesssim 45 \; \mathrm{m \; s^{-2}}$. We then find that,
\begin{equation}
    d > \frac{3 \rho_{air}(z) \times (15.5 \mathrm{\; km \; s^{-1}})^2}{4 \rho_{obj} \times (45 \mathrm{\; m \; s^{-2}})}.
\end{equation}

Adopting $\rho_{obj} \sim$3500 $\mathrm{\; kg \; m^{-3}}$, and~calculating the mean density of air over the~\citet{Shober2019} meteor's 90 s trajectory to be $\rho_{air}(z) = 3 \times 10^{-4} \mathrm{\; kg \; m^{-3}}$, the~constraint becomes,
\begin{equation}
    d \gtrsim 30 \; \mathrm{cm}.
\end{equation}

This implies that objects with diameters smaller than $30 \; \mathrm{cm}$, or~the size of the~\citet{Shober2019} meteor, slowed significantly by friction, although~this estimate depends on the shape and mean density to within a factor of a few. We therefore do not consider objects smaller than $30 \; \mathrm{cm}$.

We estimate the total number of transporting objects over the Earth's lifetime as the aforementioned values, calibrated by the~\citet{Shober2019} meteor is $\sim$6 $\times 10^{10}$, with~95\% Poisson bounds of $2 \times 10^9$--$3 \times 10^{11}$.

\section{Number of Captured~Objects}

We assume that most objects are ejected at near-zero speeds from the Solar System, mirroring the Sun's motion through the Local Standard of Rest (LSR), $(v_U^{\odot}, v_V^{\odot}, v_W^{\odot}) = (10, 11, 7) \mathrm{\;km\;s^{-1}}$ \cite{Bland-Hawthorn2016}. We adopt the following three-dimensional velocity dispersions for local stars, each corresponding to the standard deviation of a Gaussian distribution about the LSR: $\sigma_U = 35 \mathrm{\;km\;s^{-1}}$, $\sigma_V = 25 \mathrm{\;km\;s^{-1}}$, and $\sigma_W = 25 \mathrm{\;km\;s^{-1}}$ \cite{Bland-Hawthorn2016}.

Tight stellar binary systems have the largest capture cross sections,  thus we exclusively consider these systems in this analysis \cite{Ginsburg2018, Valtonen1983}. We apply the scalings in~\cite{Valtonen1982, Valtonen1983} to estimate the capture cross section of a stellar binary system of two solar-mass stars (such as Alpha Centauri A  and  B) as a function of relative speed of the system and the object, assuming that the orbital speed of the bound orbit is near zero,
\begin{equation}
    A_{c} \sim
  \begin{cases}
    2 \times 10^8 \mathrm{\; AU^2} \;a_{\alpha}^{-1} \; v_0^{-7} , & \text{$v_0 \geq 0.87,$} \\
    2 \times 10^8 \mathrm{\; AU^2}\; a_{\alpha}^{-1}\\
    \left( 10 \; v_0^{-2} \; \mathrm{ln} \left( \frac{0.87}{v_0}\right) + 2.65 \right), & \text{$v_0 \leq 0.87,$} \\
  \end{cases}
\end{equation}
where $v_0$ is the relative encounter speed in units of $\mathrm{km \; s^{-1}}$ and $a_\alpha = 23 \; \mathrm{AU}$ is the semi-major axis of this binary~system.

The local number density of solar-type stars is estimated to be $0.016 \; \mathrm{pc^{-3}}$ \cite{Bovy2017}. About 20\% of solar-type stars are in equal-mass binary systems with $a < 10 \; \mathrm{AU}$ \cite{Moe2017, Moe2019},  thus we estimate the number density of solar-type close binaries to be $n_{\star} \sim$1.6 $\times 10^{-3} \; \mathrm{pc}^{-3}$. The~solar-type binary period distribution is approximately log-normal, peaking at log(P/day) = 4.8 with dispersion of 2.3 orders of magnitude \cite{Duquennoy1991, Raghavan2010, Tokovinin2014}. Given our assumption that ejected objects mirror the Sun's speed relative to the LSR, we estimate that each object encounters a binary system every $\sim$5 $\times 10^5 \mathrm{\; years}$.

\section{Monte Carlo~Simulation}

\begin{figure}[H]
  \centering
  \includegraphics[width=0.7\linewidth]{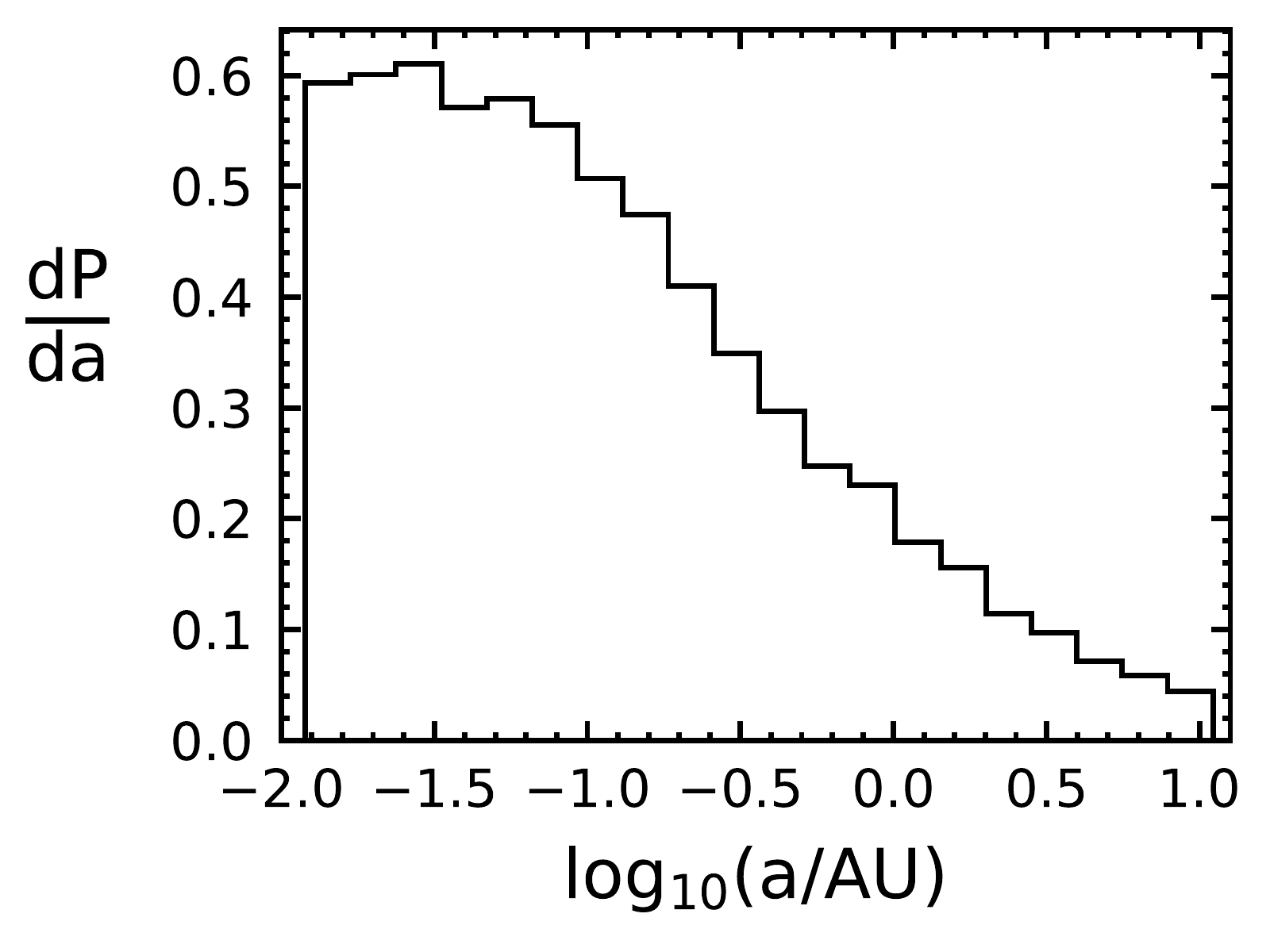}
    \caption{Distribution of semi-major axes of capturing binary systems for $1.9 \times 10^4$ simulated captured objects. P is probability and a is the semi-major axis of the binary. 
}
    \label{fig:fig1}
\end{figure}

\begin{figure}[H]
  \centering
  \includegraphics[width=0.7\linewidth]{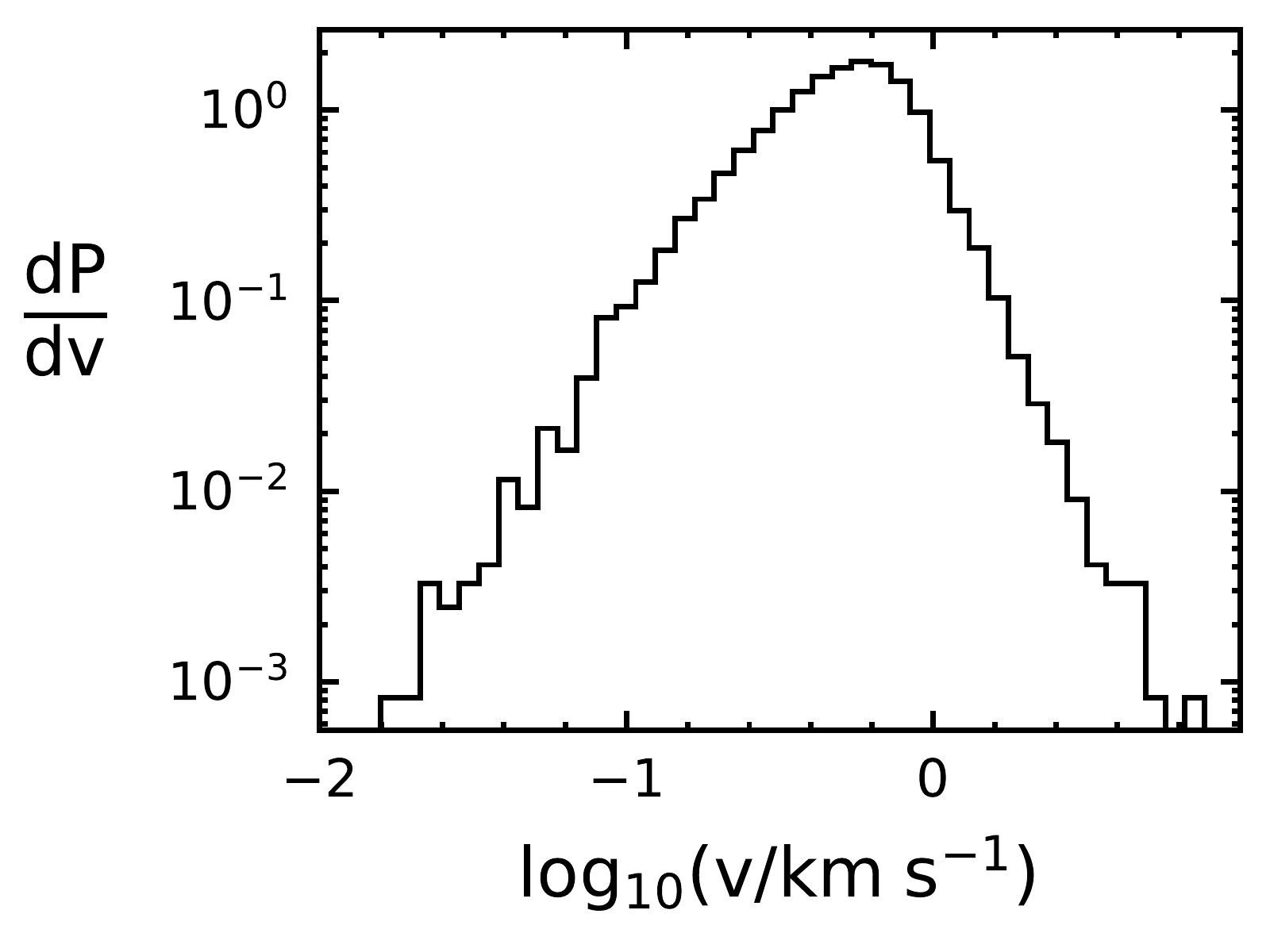}
    \caption{Distribution of relative speed at capture between $1.9 \times 10^4$ simulated captured objects and their respective binary systems. P is probability and v is the relative speed at capture.
}
    \label{fig:fig2}
\end{figure}

We use the following Monte Carlo method to determine the capture rate of life-bearing Solar System bodies. First, we draw randomly from the Gaussian distributions of local stars, centered around the LSR, to~determine the relative speed of the object with the binary system it encounters. Next, we draw the binary period from the aforementioned probability distribution, and,~assuming a circular orbit, we use Kepler's third law to translate the period to a semi-major axis length. We use the speed and semi-major axis to determine the capture cross section of the system, and~then use the capture cross section and the assumption of $n_{\star}$ to estimate the capture probability, $p_c$. We apply the probability $p_c$ to determine whether or not the object is captured. If~the object is captured, we record the time of capture, the~relative speed, and~the semi-major axis of the binary system. If~the object is not captured, we step to the next encounter and repeat the process. We ran our code for $10^9$ particles over $1.5 \times 10^7$~years, resulting in a capture probability of $6.4 \times 10^{-7}$ per encounter. The~distributions of $v$ and $a$ for the captured objects are shown in Figures~\ref{fig:fig1} and \ref{fig:fig2}, indicating that binaries with smaller semi-major lead to higher incidences of capture and that the relative speed that leads to the highest chance of capture is  within  $0.1$--$1 \mathrm{\; km \; s^{-1}}$.

Significant slow-down for centimeter-scale objects of density $\sim$3500 $\; \mathrm{kg \; m^{-3}}$ travelling through the ISM occurs when the accumulated ISM mass is comparable to the mass of the object \cite{Bialy2018},
\begin{equation}
    1.4 m_p n_p L A \sim m_i \; \; ,
\end{equation}
where $m_p$ is the proton mass; $n_p$ is the mean proton number density of the ISM, taken to be $1 \; \mathrm{cm^{-3}}$; L is the distance travelled; A is the cross-sectional area of the object;  and~$m_i$ is the mass of the object. For~the aforementioned values, we find that the timescale on which significant slow-down occurs is several times longer than the age of the Universe. We therefore disregard slow-down~effects.

Over 3 Gyr, an~order unity assumption for the timescale on which life has likely existed in the atmosphere, the~single-encounter capture probability of $\sim$6.4 $\times 10^{-7}$ leads to a total capture probability of $\sim$4 $\times 10^{-3}$, or~$\sim$2 $\times 10^{8}$ objects, with~95\% Poisson bounds of $10^{7}$--$10^{9}$. By~integrating the expression for the surviving proportion of polyextremophile colonies over a time $t$, $\exp{(-t/ 10^5 \mathrm{\; years})}$, multiplied by the constant single-encounter capture rate, we find that $\sim$4 $\times 10^{-9}$ of all objects were captured and hosted a living polyextremophile colony at the time of capture, assuming each object obtained one  polyextremophile colony after interacting with the Earth's atmosphere. We therefore estimate that $\sim$200 potentially life-bearing Solar System bodies could have been captured by exoplanetary systems over the lifetime of the Solar System while there still may have possibly been living microbes on the object, with~95\% Poisson bounds of $10$--$1000$ objects.

\unskip

\section{Conclusions}
We calculated the likelihood of potentially life-bearing Solar System bodies to get captured by exoplanetary systems. The~total number of objects captured by exoplanetary systems over the lifetime of the Solar System is $10^{7}$--$10^{9}$, with~the total number of objects with the possibility of living microbes on them at the time of capture estimated to be $10$--$1000$.

Panspermia is a multi-stage process and not all steps are addressed here, most notably the delivery of microbes to a planet within    binary systems. Many of the probability estimates derived here are related to explicitly stated fiducial values, and~are easily scaled to accommodate future research. Further detections of Earthgrazing objects transferred into Jupiter-crossing orbits will allow for more precise estimates of this process. Measurements of the abundance of polyextremophiles in the atmosphere will also refine the estimates made in this paper.  Additional studies of polyextremophiles in space will advance the biological understanding of this panspermia~channel.

\section*{Acknowledgements}
This work was supported in part by a grant from the Breakthrough Prize Foundation. 

\end{document}